\documentclass[12pt]{article}
\usepackage{draft,graphicx}
\usepackage{dcolumn}
\usepackage{bm,tikz-cd}

\def\ma{\mathcal}
\def\ie{\begin{equation}\begin{aligned}}
\def\fe{\end{aligned}\end{equation}}

\begin{document}

\title{The Aharony-Bergman-Jafferis-Maldacena theory on a circle}

\author{Yi-Xiao Tao}
\email{taoyx21@mails.tsinghua.edu.cn}
\address{Department of Mathematical Sciences, Tsinghua University, Beijing 100084, China
}%

\begin{abstract}
In this work, we bootstrap the 4-point correlators on the 1D celestial circle using 3D symmetries in the Aharony-Bergman-Jafferis-Maldacena theory as constraints. We find that the dual inversion property is strong enough to replace the crossing symmetry condition (or cyclic invariant condition) when bootstrapping. We also give some results about the conformal block expansion coefficients which contain the spectrum. Furthermore, we extract the OPE spectrum from the multi-collinear limit since all 3-point ABJM amplitudes vanish. Although we studied a specific theory, the methods used are valid for more general cases. 
\end{abstract}
\newpage
\tableofcontents
\newpage
\section{Introduction and review}
The Aharony-Bergman-Jafferis-Maldacena theory (ABJM) is a 3D Chern-Simons matter theory with $\mathcal{N}=6$ superconformal symmetries. Since the similarity to the $\mathcal{N}=4$ super Yang-Mills theory (SYM) and the applications in AdS/CFT \cite{Aharony:2008ug}, people have studied ABJM for a long time and have obtained many remarkable results, both from the CFT aspect and the amplitude aspect \cite{Benna:2008zy,Chen:2011vv,Drukker:2010nc,Drukker:2011zy,Gang:2010gy,He:2022lfz,He:2023rou,Hosomichi:2008jb,Huang:2010qy}. The matter fields in ABJM can be figured out from the R-symmetry $SO(6)=SU(4)$, including 4 complex scalars $X_\mathsf{A}$ and four complex fermions $\Psi^{\mathsf{A}a}$ as well as their conjugate \cite{Elvang:2013cua}. They transform in the fundamental (or anti-fundamental) representation of $SU(4)$ and $\mathsf{A}=1,2,3,4$. We can introduce 3 anti-commuting variables $\eta^A$ to arrange these states in on-shell superspace, namely
\ie
\Phi&=X^4+\eta^A\psi_A-\frac{1}{2}\epsilon_{ABC}\eta^A\eta^BX^C-\eta^1\eta^2\eta^3\psi_4\\
\bar{\Psi}&=\bar{\psi}^4+\eta^A\bar{X}_A-\frac{1}{2}\epsilon_{ABC}\eta^A\eta^B\bar{\psi}^C-\eta^1\eta^2\eta^3\bar{X}_4,
\fe
where we have split $\mathsf{A}$ to $\{A,4\}$. The superamplitudes involving these two superfields are constrained by superconformal symmetries. The amplitudes of the matter fields can be obtained by selecting the components of the anti-commuting variables in the superamplitudes and are hence called ``partial amplitudes". ABJM has many good properties and we will see them later. 

Recently, celestial holography, which translates the 4D amplitudes in the bulk to CFT correlators on the celestial sphere that is equivalent to the lightlike infinity of the origin \cite{Pasterski:2016qvg,Pasterski:2017kqt}, has given us many illustrating results \cite{Pasterski:2017ylz,Donnay:2018neh,Pate:2019lpp,Pate:2019mfs,Chang:2021wvv}. Inspired by AdS/CFT, celestial holography aims to help us understand quantum gravity. In addition, it also helps us understand other theories from a new aspect. For massless amplitudes, this correspondence can be realized by the Mellin transformation on the magnitude of the external momenta, namely (for 4D amplitudes)
\ie
\ma{A}_n=\int_{0}^{\infty}\prod (d\omega_i\omega_i^{\Delta_i-1})A_n(p_i=\epsilon_i\omega_i\hat{p}_i)\delta^{4}(\sum_ip_i)
\fe
where $\ma{A}_n$ and $A_n$ are celestial amplitudes and 4D amplitudes respectively. Note that $\hat{p}_i$ is parameterized by the coordinates of the celestial sphere and $\epsilon=\pm1$ for outgoing and incoming. The celestial holography can be easily generalized to 3D by using the following parameterization \cite{Lam:2017ofc}:
\ie\label{para}
p_i=\epsilon_i\omega_i(1+x_i^2,2x_i,1-x_i^2).
\fe
Then after the Mellin transformation, we will obtain the CFT correlators on the celestial circle from the 3D amplitudes. To study the spectrum of such celestial CFT (CCFT) is still a significant problem.

One may ask if we can obtain some new perspectives on the ABJM via CCFT. Furthermore, can we bootstrap the CCFT corresponding to ABJM? In this paper, we will answer these two questions tentatively. We will first show how to bootstrap the 4-point correlators of the CCFT using the 3D symmetry as constraints and then show the conformal block expansion of the 4-point correlator and the loop correction. Finally, we will show the multi-collinear behavior \cite{Ball:2023sdz,Guevara:2024ixn} and the in-in (or out-out) OPE of this CCFT. We will see that one can obtain some important information of the CCFT corresponding to ABJM, like the spectrum, from a total CFT perspective. 

This paper is organized as follows. In section \ref{sec2}, we will use the celestial version of 3D spin-helicity formalism, which was proposed in \cite{Tao:2023wls}, to express the 3D symmetries as 1D constraints for tree-level celestial amplitudes\footnote{As a supplement, for the celestial version of 4D massive spinor-helicity formalism, see \cite{Chattopadhyay:2024kdq}.}. We will also write down the ansatz for the 4-point tree-level celestial amplitudes. In section \ref{sec3}, we will solve these constraints. In section \ref{sec4}, we will first introduce how the dual inversion acts on the celestial variables for 4-point amplitudes. In section \ref{sec5}, we will investigate the spectrum of this CCFT and the influence of the loop corrections. In section \ref{sec6}, we will consider the multi-collinear limit to obtain the spectrum of the celestial OPE.

\section{3D symmetries $\longleftrightarrow$ 1D constraints}\label{sec2}
To translate the 3D symmetries to 1D constraints, it is useful to consider the 3D spin-helicity formalism, which comes from the dimensional reduction of the 4D case \cite{Elvang:2013cua,Tao:2023wls}:
\ie
\lambda_{a}=\sqrt{-2\epsilon\omega}\left(\begin{array}{c} x\\-1  \end{array}\right)\\
\lambda^{a}=\varepsilon^{ab}\lambda_{b}=-\sqrt{-2\epsilon\omega}\left(\begin{array}{c} 1\\x  \end{array}\right)
\fe
with $p^{\mu}_{ab}=\lambda_a\lambda_b$ and the parameterization \ref{para}. From the chain rules, the derivatives of these variables are
\ie
\frac{\partial}{\partial \lambda^{1}}=\frac{1}{\sqrt{-2\epsilon\omega}}(x\frac{\partial}{\partial x}-2\omega\frac{\partial}{\partial\omega})\\
\frac{\partial}{\partial \lambda^{2}}=-\frac{1}{\sqrt{-2\epsilon\omega}}\frac{\partial}{\partial x}
\fe
Since 1D global conformal symmetry is the same as the 3D Lorentz symmetry, the 1D constraints will only come from the translation operators $P^{ab}$, the dilatation operator $D$, the $U(1)$ R-symmetry operator $R$, and the supercharge $Q^{aA}$:
\ie
&P^{ab}=\sum_i\lambda_i^a\lambda_i^b, D=\sum_i(\lambda_i^a\partial_{\lambda^a_i}-\frac{1}{2})=\sum_i(\omega\frac{\partial}{\partial\omega}+\frac{1}{2}), \\
&R\equiv R^{\;\;C}_C=\sum_i(\eta_{i}^C\partial_{\eta_{i}^C}-3/2), Q^{aA}=\sum_i\lambda^a_i\partial_{\eta_{i}^A}
\fe
From the definition of celestial amplitudes, the celestial version of these operators are
\ie
P^{ab}\ma{A}_n&=0 \longrightarrow \sum_i\epsilon_ix_i^j\ma{T}_i\ma{A}_n=0, \text{ for $j=0,1,2$.}\\
D\ma{A}_n&=(\frac{n}{2}-\sum_i\Delta_i)\ma{A}_n=0\\
R\ma{A}_n&=\sum_i(\eta_{i}^C\partial_{\eta_{i}^C}-3/2)\ma{A}_n=0\\
Q^{aA}\ma{A}_n&=0\longrightarrow-\sum_i\sqrt{-2\epsilon_i}\ma{T}_i^{1/2}\left(\begin{array}{c} 1\\x_i  \end{array}\right)\partial_{\eta_{i}^{A}}\ma{A}_n=0.
\fe
where the operator $\ma{T}_i$ can add 1 to the $\Delta_i$ in the 1D correlator $\ma{A}_n$ and the label $j$ is the power of $x_i$. In ABJM, besides the usual superconformal symmetry, one can also find the dual superconformal symmetry \cite{Huang:2010qy} which will be manifest after using the dual coordinates. Such dual superconformal symmetry can generate a larger algebra together with the original superconformal symmetry, which is a Yangian \cite{Drummond:2009fd}. This fact is similar to the case in the $\ma{N}=4$ SYM, and we only need to consider two level-1 operators in the Yangian. When acting on the 1D correlators, they are equivalent to the dual conformal boost operator and its superpartner. The dual conformal boost operator for ABJM amplitudes reads \cite{Huang:2010qy}
\ie
\ma{K}^{ab}\ma{A}_n=&-\frac{1}{4}\sum_{k<i}\bigg[\lambda_k^{(a}\lambda_i^{b)}\bigg(\lambda_k^{\gamma}\frac{\partial}{\partial \lambda_i^{\gamma}}-\lambda_i^{\gamma}\frac{\partial}{\partial \lambda_k^{\gamma}}\\
&+\eta_k^{B}\frac{\partial}{\partial \eta_i^{B}}-\eta_i^{B}\frac{\partial}{\partial \eta_k^{B}}\bigg)+2\lambda_k^a\lambda_k^b\bigg].
\fe
Such Yangian symmetry has some new symmetries compared with the original one and we will later see that it can replace the crossing symmetry in the following ansatz. However, let us now ignore the dual conformal symmetry first. We conclude our ansatz for 4-point tree correlators here:
\begin{enumerate}
    \item It has the form $$
    \ma{A}_4=g(\epsilon_i,\Delta_i)f(z,\eta_i^A)\prod_{i<j}x_{ij}^{\Delta/3-\Delta_i-\Delta_j}$$ with $z=\frac{x_{12}x_{34}}{x_{13}x_{24}}$ the cross-ratio and $\Delta=\sum_{i=1}^4\Delta_i$.
    \item The operator $P^{ab},D,R,Q^{aA}$ will annihilate the correlator.
    \item The function $f(z)$ should have the form $f(z)\propto z^m(1-z)^n$, which ensures that this is a tree-level correlator.
    \item Crossing symmetry under $1\leftrightarrow3$.
    \item Modified crossing symmetry under $2\leftrightarrow3$
\end{enumerate}
The first ansatz comes from the 1D conformal symmetry, and the third one will forbid factors involving logarithm terms. The ansatz for $f(z)$ is necessary since when we take $z\to 1-z$ and $z\to 1/z$ we will obtain an overall factor $z^a(1-z)^b$ with some $a$, $b$ from the crossing symmetry. The last two ansatz come from the usual crossing symmetry for 4-point correlators. The crossing symmetry part can be understood as the cyclic invariant condition. There will be an extra factor when we exchange labels 2 and 3 compared with the usual crossing symmetry since this case will correspond to the channel $s\leftrightarrow u$. We will see this point later.

\section{Four-point correlators from 1D constraints}\label{sec3}
Although it is well-known how to obtain the 4-point ABJM amplitudes at the tree level directly from the symmetry, we still need to figure out how these constraints help us to construct the 4-point 1D correlators. The translation operators will force
\ie
g(\epsilon_i,\Delta_i)&\propto\prod_{i=1}^4((-1)^{i+1}\epsilon_1\epsilon_i)^{\Delta_i},
\fe
while the $U(1)$ R-symmetry operator will force the degree of freedom of anti-commuting variables to be 6. This constraint also forbids the appearance of odd-point correlators. The supersymmetry constraints can be solved using momentum conservation, i.e. we should have
\ie
\partial_{\eta_i^A}\ma{A}_4\sim \sqrt{\epsilon_i\ma{T}_i}\ma{A}_4 \ \ \text{or}\ \ \partial_{\eta_i^A}\ma{A}_4\sim \sqrt{\epsilon_i\ma{T}_i}x_i\ma{A}_4
\fe
Then we can solve $f(z)$ as an operator:
\ie
f(z,\eta_i^A)=h(z) \prod_{A=1}^3(\sum_{i<j}\sqrt{\epsilon_i\epsilon_j\ma{T}_i\ma{T}_j}x_{ij}\eta_i^A\eta_j^A)
\fe
with $h(z)$ a function of $z$, which should have the form $h(z)=z^a(1-z)^b$ and can be figured out from the formalism of $f(z)$ in the ansatz. The factor $x_{ij}$ is introduced in order to let $f(z)$ only dependent with $z$ after acting the operator $\sqrt{\ma{T}_i\ma{T}_j}$ on the following factor. Note that in our case the crossing symmetry is only valid for the total correlator rather than the partial correlators since the anticommuting variables also need to transform when we consider the crossing symmetry. After writing the correlator in the following form assuming that all $\Delta_i$ are the same:
\ie
\ma{A}_4=\frac{c(z,\eta_i)}{x_{12}^{\Delta/2}x_{34}^{\Delta/2}},
\fe
the crossing symmetry reads \cite{Lam:2017ofc}:
\ie
&c(1-z,\eta_3,\eta_2,\eta_1,\eta_4)\sim(\frac{1-z}{z})^{\Delta/2}c(1-z,\eta_1,\eta_2,\eta_3,\eta_4)\\
&c(z,\eta_1,\eta_2,\eta_3,\eta_4)\sim z^{\Delta/2+(2a+b)}c(1/z,\eta_1,\eta_3,\eta_2,\eta_4)
\fe
Here $\sim$ means ``up to some possible minus signs". The first equation is for $1\leftrightarrow3$ and the second one is for $2\leftrightarrow3$. The factor $z^{2a+b}$ comes from $h(1/z)\sim z^{-2a-b}h(z)$. The solution is
\ie
h(z)=z^{-1/6}(1-z)^{-1/6},
\fe
Collecting all the elements, we will obtain the 4-point correlator
\ie\label{4pt}
\ma{A}_4=&\delta(\sum_{i=1}^4\Delta_i-2)z^{-1/6}(1-z)^{-1/6}\prod_{A=1}^3(\sum_{i<j}\sqrt{\epsilon_i\epsilon_j\ma{T}_i\ma{T}_j}x_{ij}\eta_i^A\eta_j^A)\circ\bigg[\prod_{i=1}^4((-1)^{i+1}\epsilon_1\epsilon_i)^{\Delta_i}\prod_{i<j}x_{ij}^{\Delta/3-\Delta_i-\Delta_j}\bigg]
\fe
where the delta function comes from the constraint from the dilatation operator. This result coincides with doing the Mellin transformation on the 4-pt color-ordered amplitude
\ie\label{4ptamp}
A_4(\bar{\Psi}\Phi\bar{\Psi}\Phi)=\frac{\delta^3(P)}{\langle14\rangle\langle43\rangle}\prod_{A=1}^{3}(\sum_{i<j}\langle ij\rangle\eta_{iA}\eta_{jA})
\fe
up to an overall coefficient independent of $\Delta_i$ \footnote{Since we only consider the 4-point amplitudes of ABJM, we will not consider the 4-index structure constant $f^{a_2a_4\bar{a}_1\bar{a}_3}$.}. In the above 4-point amplitude, it is clear that the cyclic invariant will cause two crossing symmetry conditions because of the relations $\langle12\rangle\sim\langle34\rangle$ and so on.

Now we show the calculation of the Mellin transformation of \eqref{4ptamp} in the following. Using the celestial coordinate, the 4-pt amplitude is
\ie
A_4[\bar{\Psi}_1\Phi_2\bar{\Psi}_3\Phi_4]=-2\frac{\delta^3(P)}{\epsilon_4\omega_4\sqrt{\epsilon_1\epsilon_3\omega_1\omega_3}x_{14}x_{43}}\prod_{A=1}^{3}(\sum_{i<j}\sqrt{\epsilon_i\epsilon_j\omega_i\omega_j}x_{ij}\eta_{iA}\eta_{jA}).
\fe
The 4-pt delta function can be parameterized as follows:
\ie\label{mc}
\delta^3(P)=\delta(\omega_2+\frac{\epsilon_1x_{13}x_{14}}{\epsilon_2x_{23}x_{24}}\omega_1)\delta(\omega_3-\frac{\epsilon_1x_{12}x_{14}}{\epsilon_3x_{23}x_{34}}\omega_1)\delta(\omega_4+\frac{\epsilon_1x_{12}x_{13}}{\epsilon_4x_{24}x_{34}}\omega_1)|\frac{1}{4\epsilon_2\epsilon_3\epsilon_4 x_{23}x_{24}x_{34}}|.
\fe
Then the 4-pt amplitude is
\ie
&A_4[\bar{\Psi}_1\Phi_2\bar{\Psi}_3\Phi_4]=-\frac{\prod_{A=1}^{3}(\sum_{i<j}\sqrt{\epsilon_i\epsilon_j\omega_i\omega_j}x_{ij}\eta_{iA}\eta_{jA})}{\epsilon_4\omega_4\sqrt{\epsilon_1\epsilon_3\omega_1\omega_3}x_{14}x_{43}}\delta(\omega_2+\frac{\epsilon_1x_{13}x_{14}}{\epsilon_2x_{23}x_{24}}\omega_1)\delta(\omega_3-\frac{\epsilon_1x_{12}x_{14}}{\epsilon_3x_{23}x_{34}}\omega_1)\\
&\delta(\omega_4+\frac{\epsilon_1x_{12}x_{13}}{\epsilon_4x_{24}x_{34}}\omega_1)|\frac{1}{2\epsilon_2\epsilon_3\epsilon_4 x_{23}x_{24}x_{34}}|.
\fe
The celestial amplitude can be obtained by doing the following integral:
\ie
&\ma{A}_4[\bar{\Psi}_1\Phi_2\bar{\Psi}_3\Phi_4]=\int d\omega \omega^{\sum\Delta_i-6}(-\frac{\epsilon_1x_{13}x_{14}}{\epsilon_2x_{23}x_{24}})^{\Delta_2-1}(\frac{\epsilon_1x_{12}x_{14}}{\epsilon_3x_{23}x_{34}})^{\Delta_3-3/2}(-\frac{\epsilon_1x_{12}x_{13}}{\epsilon_4x_{24}x_{34}})^{\Delta_4-2}\\
&\frac{\prod_{A=1}^{3}(\sum_{i<j}\sqrt{\epsilon_i\epsilon_j\omega_i\omega_j}x_{ij}\eta_{iA}\eta_{jA})}{\epsilon_4\sqrt{\epsilon_1\epsilon_3}x_{14}x_{34}}|\frac{1}{2\epsilon_2\epsilon_3\epsilon_4 x_{23}x_{24}x_{34}}|\\
&=\delta(\sum_{i=1}^4\Delta_i-2)(\Omega_2)^{\Delta_2-1}(\Omega_3)^{\Delta_3-3/2}(\Omega_4)^{\Delta_4-2}\frac{\prod_{A=1}^{3}(\sum_{i<j}\Omega_i^{1/2}\Omega_j^{1/2}x_{ij}\eta_{iA}\eta_{jA})}{\epsilon_4\sqrt{\epsilon_1\epsilon_3}x_{14}x_{34}}|\frac{1}{2\epsilon_2\epsilon_3\epsilon_4 x_{23}x_{24}x_{34}}|
\fe
where $\Omega_1=1$, $\Omega_2=-\frac{\epsilon_1x_{13}x_{14}}{\epsilon_2x_{23}x_{24}}$, $\Omega_3=\frac{\epsilon_1x_{12}x_{14}}{\epsilon_3x_{23}x_{34}}$, $\Omega_4=-\frac{\epsilon_1x_{12}x_{13}}{\epsilon_4x_{24}x_{34}}$. One can check that this result coincide with \eqref{4pt} by comparing each partial amplitude.

One can use the celestial BCFW or other celestial recursion methods to obtain higher-point results, see \cite{Britto:2004ap,Britto:2005fq,Hu:2022bpa,Tao:2023wls}. In principle, we can obtain higher-point correlators using these methods.

\section{Dual inversion}\label{sec4}
The crossing symmetry conditions for the 4-point correlators are slightly messy, although they come from the cyclic invariant. In this section, we will use the property of dual inversion to replace them. We will also find out $h(z)$ only from the behavior under the dual inversion. 

If we write the momenta $p_i$ using the region momenta $y_i$ as $y_i-y_{i+1}=p_i$, the dual inversion will be defined as
\ie
\ma{I}(y_i^{\mu})=\frac{y_i^{\mu}}{y^2_i}.
\fe
Also for the anti-commuting variables:
\ie
|\theta_{i,A}\rangle-|\theta_{i+1,A}\rangle=|i\rangle\eta_{iA},\ \ \ma{I}(|\theta_{i,A}\rangle^{a})=|\theta_{i,A}\rangle_b\frac{y_i^{ba}}{y_i^2}.
\fe
The behavior of ABJM amplitudes under the dual inversion is
\ie
\ma{I}(A_4(\bar{\Psi}\Phi\bar{\Psi}\Phi))=\sqrt{y_1^2y_2^2y_3^2y_4^2}A_4(\bar{\Psi}\Phi\bar{\Psi}\Phi),
\fe
which will lead to the following equation for the celestial ABJM amplitudes:
\ie\label{dual}
\ma{I}(\ma{A}_4)=\sqrt{y_1^2y_2^2y_3^2y_4^2}\prod_{i=1}^{4}f_i^{\Delta_i}\ma{A}_4
\fe
with $\ma{I}(\omega_i)=f_i\omega_i$\footnote{After some calculations we can obtain $f_i=\frac{y_i^{11}y_{i+1}^2-y_{i+1}^{11}y_i^2}{y_i^2y_{i+1}^2(y_i^{11}-y_{i+1}^{11})}$. But we do not need this explicit form.}. We will show how \eqref{dual} leads to the correct answer. Let us start with finding out the behavior of $x_{ij}$ under the dual inversion. From the formula
\ie
\ma{I}(\langle i,i+1\rangle)=\frac{\langle i,i+1\rangle}{\sqrt{y_i^2y_{i+2}^2}}\to\ma{I}(\sqrt{\omega_i\omega_{i+1}}x_{i,i+1})=\frac{\sqrt{\omega_i\omega_{i+1}}x_{i,i+1}}{\sqrt{y_i^2y_{i+2}^2}},
\fe
we have the following equations:
\ie\label{dual2}
&\ma{I}(x_{12})=\frac{x_{12}}{\sqrt{y_1^2y_3^2f_1f_2}},\ma{I}(x_{23})=\frac{x_{23}}{\sqrt{y_2^2y_4^2f_2f_3}}\\
&\ma{I}(x_{34})=\frac{x_{34}}{\sqrt{y_3^2y_1^2f_3f_4}},\ma{I}(x_{14})=\frac{x_{14}}{\sqrt{y_2^2y_4^2f_1f_4}}\\
&\ma{I}(x_{24})=\frac{x_{24}}{\sqrt{f_2f_4}}\sqrt{\frac{y_2^2y_3^2y_4^2}{y_1^{10}}},\ma{I}(x_{13})=\frac{x_{13}}{\sqrt{f_1f_3}}\sqrt{\frac{y_2^2y_3^2y_4^2}{y_1^{10}}}.
\fe
We have also used the consistency of the dual inversion and the momentum conservation \eqref{mc}, and the formula $\ma{I}(\delta^{3}(\sum_{i=1}^4p_i))=y_1^6\delta^{3}(\sum_{i=1}^4p_i)$. Then we have
\ie
\ma{I}(z)=\frac{y_1^8}{y_2^2y_3^4y_4^2}z,\ma{I}(1-z)=\frac{y_1^{10}}{y_2^4y_3^2y_4^4}(1-z).
\fe

Assuming that the operator
\ie
\prod_{A=1}^3(\sum_{i<j}\sqrt{\epsilon_i\epsilon_j\ma{T}_i\ma{T}_j}x_{ij}\eta_i^A\eta_j^A)
\fe
acting on $\prod_{i<j}x_{ij}^{\Delta/3-\Delta_i-\Delta_j}$ will give an factor
\ie
\prod_{A=1}^3(\sum_{i<j}\sqrt{\epsilon_i\epsilon_j}\Omega_{ij}x_{ij}\eta_i^A\eta_j^A),
\fe
where
\ie
\Omega_{ij}=\frac{\sqrt{\ma{T}_i\ma{T}_j}\bigg(\prod_{k<l}x_{kl}^{\Delta/3-\Delta_k-\Delta_l}\bigg)}{\prod_{k<l}x_{kl}^{\Delta/3-\Delta_k-\Delta_l}}
\fe
Then such a factor will be invariant under the dual inversion after some calculations according to 
\ie\label{dual3}
\ma{I}[\prod_{A=1}^{3}(\sum_{i<j}\langle ij\rangle\eta_{iA}\eta_{jA})]=y_1^{-6}\prod_{A=1}^{3}(\sum_{i<j}\langle ij\rangle\eta_{iA}\eta_{jA}).
\fe
The factor
\ie
\prod_{i<j}x_{ij}^{\Delta/3-\Delta_i-\Delta_j}
\fe
will give a $y_1^{2\Delta}\prod_{i=1}^{4}f_i^{\Delta_i}$. Substituting the delta function $\delta(\sum_{i=1}^4\Delta_i-2)$, We find that the undetermined $h(z)$ must be
\ie
z^{-1/6}(1-z)^{-1/6}
\fe
up to a constant to obtain \eqref{dual}. Hence for a 1D CFT, we can impose the constraints \eqref{dual} and also define the dual inversion \eqref{dual2} and \eqref{dual3} for $x_{ij}$ and $\eta_i^A$ respectively. This constraint, together with the first 3 ansatz in section \ref{sec2}, will determine the 4-point correlator of the 1D CCFT dual to the 3D ABJM theory. Note that the properties \eqref{dual2} and \eqref{dual3} are valid for 4-point correlators in any 1D CFT  with the dual conformal symmetry.

\section{Conformal block expansion and loop corrections}\label{sec5}
In this section we will show the in-out OPE by computing the conformal block expansion. Using the inversion formula \cite{Atanasov:2021cje,Rutter:2020vpw}, we can obtain the coefficients of the $t$-channel conformal block expansion with $\epsilon_1=\epsilon_2=-\epsilon_3=-\epsilon_4=-1$, which corresponds to the OPE between an incoming particle and an outgoing particle. First, we rewrite the 4-point correlator in the following form:
\ie
\ma{A}_4=\delta(\sum_i\Delta_i-2)\frac{(\frac{x_{34}}{x_{14}})^{\Delta_1-\Delta_3}(\frac{x_{14}}{x_{12}})^{\Delta_2-\Delta_4}}{x_{13}^{\Delta_1+\Delta_3}x_{24}^{\Delta_2+\Delta_4}}\ma{G}(\chi)
\fe
where $\chi=1/z\in(0,1)$. We have omitted the overall factor $\prod_{i=1}^4((-1)^{i+1}\epsilon_1\epsilon_i)^{\Delta_i}$ since we only focus on the spectrum. The function $\ma{G}(\chi)$ has a general form $\chi^a(1-\chi)^b$, and the corresponding coefficient of the conformal block expansion is shown below, following the calculations in \cite{Chang:2023ttm}
\ie
\ma{C}_{a+n}^{\Delta_i}=&\frac{(a+\Delta_{13})_n(a+\Delta_{24})_n}{n!(2a+n-1)_n}{}_3F_2(\begin{array}{c}-n,2a+n-1,a+b+\Delta_{24}\\\Delta_{13}+a,\Delta_{24}+a\end{array};1)
\fe
where $n\geq0$ and the exchanging operators have weights $a+n$. We list all the pairs $(a,b)$ appearing in different partial correlators here:
\ie
&(1/2,1),(2,-1/2),(1/2,-1/2),(1,1/2),(1/2,1/2),\\
&(3/2,0),(1/2,0),(3/2,-1/2),(1,-1/2),(1,0)
\fe
The corresponding coefficients contain some extra minus sign or $i$ due to the sign of $\epsilon_i$. These coefficients also receive the loop correction. In ABJM, the 4-point planar loop amplitudes start with the 2-loop level because of the unitarity and the vanishing of 3-point tree amplitudes and it is proportional to the tree amplitude \cite{Chen:2011vv}. The loop correction operator is \footnote{The loop correction is no longer dual conformal covariant. However, it can be fixed as in the $\ma{N}=4$ SYM case \cite{Drummond:2008vq}.}
\ie
\ma{L}^{\text{2-loop}}=&\frac{-(-4\frac{x_{13}x_{14}}{x_{23}x_{24}}x_{12}^2)^{-\epsilon}-(-4\frac{x_{13}x_{14}}{x_{24}}\frac{x_{12}x_{14}}{x_{34}})^{-\epsilon}}{(2\epsilon)^2}\mu^{2\epsilon}\ma{T}_1^{-2\epsilon}\\
&+\frac{1}{2}\text{ln}^2(\frac{x_{12}x_{34}}{x_{14}x_{23}})+4\zeta(2)-3\text{ln}^22+\ma{O}(\epsilon)
\fe
Here $\mu$ is the regularization scale and $\zeta(2)=\pi^2/6$ is the Riemann zeta function $\zeta(z)$. After acting this operator on the 4-point correlator $\ma{A}_4$ we will obtain the loop correction. Such a logarithm term, which is the finite part of the loop correction, will not change the poles of the coefficients of the conformal partial wave expansion but will only change the value of the residue at these poles. The corrected coefficients can only be solved numerically and we will not show it here.

\section{Multi-collinear limit and the OPE}\label{sec6}
In this section we will compute in-in (or out-out) OPE by the multi-collinear limit. Usually, one can obtain such OPE by collinear limit. However, since odd-point correlators vanish, the collinear limit fails to find the OPE information in our case. Fortunately, it is possible to consider the multi-collinear limit. We will focus on this case and find the information about the OPE of two outgoing particles.

Following \cite{Ball:2023sdz}, We can generalize the consecutive OPE in 2D CCFT to the 1D case. Considering a $n$-pt correlator in our case and assuming $\epsilon_1=\epsilon_2=\epsilon_3=1$, we set
\ie
x_1=x_3+\epsilon, x_2=x_3+\delta\epsilon.
\fe
Consequently this will cause
\ie
&\langle13\rangle=-2\sqrt{\omega_1\omega_3}\epsilon\\
&\langle23\rangle=-2\sqrt{\omega_2\omega_3}\delta\epsilon\\
&\langle12\rangle=-2\sqrt{\omega_1\omega_2}\epsilon(1-\delta)
\fe
From now on we will consider the partial correlator $\ma{A}_{n}(\bar{X}X\bar{X}X\cdots)$ without loss of generality, where $X=X^4$ and $\bar{X}=\bar{X}_4$. The partial fields in $\cdots$ are arbitrary. We first factorize the $n$-point amplitudes as follows:
\ie
A^{\text{factorized}}_n=\int d^3\eta A_4\frac{1}{p_{123}^2}A_{n-3}=\int d^3\eta\frac{\delta^3(P)}{\langle12\rangle\langle23\rangle}\prod_{A=1}^{3}(\sum_{i<j}\langle ij\rangle\eta_{iA}\eta_{jA})\frac{1}{(\langle12\rangle^2+\langle13\rangle^2+\langle23\rangle^2)}A_{n-3}
\fe
where $\eta$ is the anti-commuting variable of the cutting leg with momentum $p_{123}$. Focusing on the $\eta_1^3\eta_3^3$ term, we have
\ie
A^{\text{factorized}}_n(\bar{X}X\bar{X}X\cdots)=-\frac{1}{2\sqrt{\omega_1\omega_2}(1-\delta)\sqrt{\omega_2\omega_3}\delta\epsilon}\frac{\omega_1\omega_3\sqrt{\omega_1\omega_3}}{\omega_1\omega_2(1-\delta)^2+\omega_1\omega_3+\omega_2\omega_3\delta^2}A_{n-3}
\fe
After changing the variables $\omega_1,\omega_2,\omega_3$ to
\ie
\omega=\omega_1+\omega_2+\omega_3,s=\frac{\omega_1}{\omega_1+\omega_2+\omega_3},t=\frac{\omega_2}{\omega_1+\omega_2+\omega_3}
\fe
We will obtain the following celestial result regardless of the higher $\epsilon$ order terms\footnote{We have also set $\epsilon=0$ in the delta function.}:
\ie
\ma{A}^{\text{factorized}}_n=&-\int_0^{\infty} d\omega \int_0^1ds\int_0^{1-s}dt\omega^{\Delta_1+\Delta_2+\Delta_3-2}s^{\Delta_1}t^{\Delta_2-2}\\
&\times\frac{(1-s-t)^{\Delta_3}}{2(1-\delta)\delta\epsilon}\frac{A_{n-3}\delta^{3}(P)}{st(1-\delta)^2+s(1-s-t)+t(1-s-t)\delta^2}=I\ma{A}_{n-3}.
\fe
with
\ie\label{prefactor}
I=-\int_0^1ds\int_0^{1-s}dt\frac{(1-s-t)^{\Delta_3}}{2(1-\delta)\delta\epsilon}\frac{s^{\Delta_1}t^{\Delta_2-2}}{st(1-\delta)^2+s(1-s-t)+t(1-s-t)\delta^2}
\fe
This formula gives the following leading multi-OPE
\ie
\bar{X}_{\Delta_1}X_{\Delta_2}\bar{X}_{\Delta_3}\sim \bar{X}_{\Delta_1+\Delta_2+\Delta_3-1}
\fe
with the multi-OPE prefactor $I$ (including the pole factor and the multi-OPE coefficient). When we set $\delta\to0,1,\infty$, we will obtain the result for the consecutive OPE. However, as mentioned before, the collinear limit between 2 momenta is not well-defined, which means that we should keep this $\delta$ to be arbitrary except $0,1,\infty$. Note that this multi-collinear limit cannot be obtained from the 6-point amplitudes directly since there are other channels besides the channel we considered in the derivation of the multi-collinear limit. Consider the case $\delta\to0$, the prefactor is 
\ie
I_{\Delta_1>0}=&\frac{-1}{2x_{23}}B(\Delta_2-1,\Delta_3+1)B(\Delta_1,\Delta_2+\Delta_3-1)\,{}_{2}F_1(1,\Delta_2-1,\Delta_2+\Delta_3,2\delta)
\fe
when $\Delta_1>0$. Here we only ignore the $\delta^2$ terms in the denominator. When $-1<\Delta_1<0$, the prefactor becomes a branch cut term:
\ie
I_{-1<\Delta_1<0}=&\frac{\pi x_{23}^{2\Delta_1-1}}{2x_{13}^{2\Delta_1}\sin{(\Delta_1\pi)}}B(\Delta_1+\Delta_2-1,\Delta_1+\Delta_3+1)\,{}_{2}F_1(\Delta_1+1,\Delta_1+\Delta_2-1,2\Delta_1+\Delta_2+\Delta_3,2\delta)
\fe
Hence the total leading multi-OPE (w.r.t $\delta$) is
\ie\label{ope}
\bar{X}_{\Delta_1}(x_1)X_{\Delta_2}(x_2)\bar{X}_{\Delta_3}(x_3)\sim&\bigg[\frac{-1}{2x_{23}}B(\Delta_2-1,\Delta_3+1)B(\Delta_1,\Delta_2+\Delta_3-1)\\
&+\frac{\pi x_{23}^{2\Delta_1-1}}{2x_{13}^{2\Delta_1}\sin{(\Delta_1\pi)}}B(\Delta_1+\Delta_2-1,\Delta_1+\Delta_3+1))\bigg]\bar{X}_{\Delta_1+\Delta_2+\Delta_3-1}(x_3)
\fe
The dominant term of these two terms depends on the value of $\Delta_1$, more precisely, the real part of $\Delta_1$. The multi-OPE at the higher order can be obtained by expanding some hypergeometric functions and also by considering the descendants of $X$. In multi-OPE, we can find out some information about the OPE. Assuming that $\Delta_1$ is in the principal series\footnote{When considering the case $-1<\Delta_1<0$, one only needs to consider the singularity $x_{23}^{2\Delta_1-1}$, which depends on $\Delta_1$. Hence, this OPE is ill-defined.}, we only need to consider $I_{\Delta_1>0}$. Since $\delta\to0$ corresponds to doing the $23$-OPE first, we find that the first term of \eqref{ope} corresponds to the following operator in $X_{\Delta_2}(x_2)\bar{X}_{\Delta_3}(x_3)$ OPE by the conformal weight analysis:
\ie
X_{\Delta_2}(x_2)\bar{X}_{\Delta_3}(x_3)\sim \frac{C}{x_{23}}\ma{O}_{\Delta_2+\Delta_3-1}(x_3)
\fe
with a constant $C$. However, all 3-point amplitudes in ABJM vanish; hence, the operator $\ma{O}$ must be a 2-particle operator \cite{Guevara:2024ixn} in this leading OPE. Note that we also have 4-point correlators with 2 bosonic fields and 2 fermionic fields. The only choice for such a primary operator is
\ie
\ma{O}_{\Delta_2+\Delta_3-1}\sim \ \text{combination of some} \ :\bar{\psi}\psi:_{\Delta_2+\Delta_3-1}
\fe
where $\psi^{\mathsf{A}}$ are the complex fermionic fields in the ABJM. This observation matches the Lagrangian of ABJM.

By expanding the hypergeometric function w.r.t $\delta$, we find that the poles in the multi-OPE are $x_{23}^{n-1}/x_{13}^n$ with $n=0,1,\cdots$ the order of $\delta$. The $1/x_{13}$ pole, which comes from the higher order of the first term, implies the OPE involving a 2-particle operator $X\bar{X}\sim :X\bar{X}:$ and $X\bar{X}\sim\partial_x\ma{O}$:
\ie
\bar{X}_{\Delta_1}(x_1)\big[C\partial\ma{O}(x_3)+:X_{\Delta_2}\bar{X}_{\Delta_3}:(x_3)\big]\sim-\frac{1}{x_{13}}B(\Delta_2,\Delta_3+1)B(\Delta_1,\Delta_2+\Delta_3-1)X_{\Delta_1+\Delta_2+\Delta_3-1}(x_3)
\fe

For higher $\delta$ orders, these poles will correspond to some descendants of $:X\bar{X}:$ and $\ma{O}$. In this case, the $\delta^2$ terms in \eqref{prefactor} cannot be ignored, and we need to expand the $\delta^2$ terms in it. However, it is still unknown how to obtain the specific OPE coefficients of the $X\bar{X}$ OPE as there are two types of operators $:XX:$ and $\ma{O}$ and their corresponding coefficients mix. We leave this to future work.

It is important to note that from this multi-OPE we find that there will not be an OPE like $X\bar{X}\sim X$ as expected, since there will not appear a $1/(x_{13}x_{23})$ pole. Then, from the multi-OPE perspective, we find the interaction information from the spectrum.

\section{Discussion}
We have demonstrated how to bootstrap the 4-point correlator in the 1D CCFT dual to the ABJM using 3D symmetries as constraints in 1D CFTs. We find that the dual inversion property for 4-point correlators, which is also valid for other 1D CFTs with dual conformal symmetries, gives a stronger constraint than the crossing symmetry (or cyclic invariant). Meanwhile, from the construction of the dual inversion, we can define some building blocks with given behaviors under the dual inversion and then construct different 4-point correlators with desired properties under the dual inversion. Furthermore, we calculated the conformal block expansion coefficients to learn more about the spectrum of this CCFT. We then studied the multi-collinear limit of this CCFT from the amplitude to the celestial version and obtained the multi-OPE. In this way, we find a special 2-particle operator in the OPE spectrum. Note that we only obtain the spectrum of OPE from the multi-OPE; the specific coefficients of each operator are still unknown since different operators will contribute to the same term in the multi-OPE. How to obtain the whole $X\bar{X}$ OPE will be left to future work.

A natural question is how to bootstrap higher-point correlators. We can use some recursion methods, but can we obtain the 6-point correlators directly using some basic properties in this 1D CCFT? To realize this, the understanding of unitarity is significant \cite{Lam:2017ofc,Garcia-Sepulveda:2022lga,Liu:2024vmx}. A more basic question is what this 1D CCFT is. We have obtained much information about this 1D CCFT, can we find a 1D CCFT with the same feature?  

There is a relation between $\ma{N}=4$ SYM amplitudes and ABJM amplitudes at the tree level from the Grassmannian aspect \cite{Huang:2021jlh,He:2021llb}. Also see \cite{Huang:2013owa} for the so-called ``orthogonal Grassmannian" which gives the ABJM tree amplitudes. Another interesting question is how to realize it on the celestial sphere. Suppose we believe celestial amplitudes will tell us some information that is hard to find on the amplitude side, it will be interesting to see whether the relation above somehow manifests on the CCFT side. A celestial version of the Grassmannian can be found in \cite{Ferro:2021dub}. And for celestial $\ma{N}=4$ SYM, see \cite{Jiang:2021xzy,Hu:2021lrx}. Moreover, at the loop level, the ABJM integrands can be obtained from the $\ma{N}=4$ SYM integrands by dimensional reduction \cite{He:2022cup}, can we realize this procedure from the celestial sphere to the celestial circle? This is a similar case to the relation between tree amplitudes as we considered before. Accordingly, we can also ask if there are some relations between these two CCFTs.

\section*{Acknowledgement}
The author would like to thank Qu Cao, Song He, Wen-Jie Ma and Walker Melton for the valuable discussion. YT is partly supported by the National Key R\&D Program of China (NO. 2020YFA0713000). YT also wants to thank ``iTHEMS-YITP Workshop: Bootstrap, Localization and Holography" and ``Strings 2024 Conference" for providing environments full of ideas.

\bibliographystyle{JHEP}

\bibliography{abjm}

\end{document}